\newif\ifmydraft\mydrafttrue
\newif\ifmyfull\myfullfalse
\newif\ifmyanon\myanonfalse
\documentclass[runningheads]{llncs}
\usepackage[T1]{fontenc}
\usepackage{graphicx}

\usepackage{amssymb}
\usepackage{amsfonts}
\usepackage{algorithmic}
\usepackage{graphicx}
\usepackage{textcomp}
\usepackage[dvipsnames]{xcolor}
\usepackage{xspace}
\usepackage{hyperref}
\usepackage{bcprules}
\usepackage{xargs}
\usepackage{siunitx}
\usepackage{float}
\usepackage{enumitem}
\usepackage[most]{tcolorbox}
\usepackage{stmaryrd}
\usepackage{subcaption}
\usepackage{multirow}
\usepackage{booktabs}
\usepackage{nth}
\usepackage{placeins}

\usepackage{tikz}
\usepackage{fontawesome5}  %
\usetikzlibrary{shapes.geometric, arrows.meta, positioning, fit, calc}

\tikzstyle{process} = [rectangle, minimum width=3.5cm, minimum height=0.8cm, text centered, draw=black, fill=blue!20]
\tikzstyle{arrow} = [thick,->,>=Stealth]
\tikzstyle{io} = [ellipse, minimum width=3.5cm, minimum height=1cm, text centered, draw=black, fill=green!20]
\tikzstyle{eprocess} = [rectangle, minimum width=3.5cm, minimum height=1cm, text centered, draw=black, fill=gray!10, line width=0.7mm, dashed]

\definecolor{boxcolor}{RGB}{220,230,241}
\definecolor{orangebox}{RGB}{255,183,77}
\definecolor{arrowcolor}{RGB}{79,129,189}
\definecolor{textcolor}{RGB}{51,51,51}

\newlist{rqlist}{enumerate}{1}
\setlist[rqlist,1]{
  label=\bf RQ\arabic*.,  %
  leftmargin=*,
  labelsep=0.5em,
  align=left,
  itemsep=1ex,         %
  topsep=1ex
}

\tcbset{
  rqanswerbox/.style={
    colback=gray!10,       %
    colframe=black,     %
    boxrule=0.3pt,
    arc=4pt,
    left=6pt,
    right=6pt,
    top=6pt,
    bottom=6pt,
    enhanced
  }
}

\usepackage{listings}
\definecolor{codegreen}{rgb}{0,0.4,0}
\definecolor{codeblue}{rgb}{0,0,0.6}
\definecolor{codegray}{rgb}{0.5,0.5,0.5}
\definecolor{codepurple}{rgb}{0.58,0,0.82}
\definecolor{backcolour}{rgb}{0.95,0.95,0.92}

\lstdefinestyle{mystyle}{
    backgroundcolor=\color{white},   
    commentstyle=\color{codegreen},
    numberstyle=\tiny,
    stringstyle=\color{codeblue},
    basicstyle=\ttfamily\footnotesize,
    breakatwhitespace=false,         
    breaklines=true,                 
    captionpos=b,                    
    keepspaces=true,                 
    numbersep=5pt,                  
    showspaces=false,                
    showstringspaces=false,
    upquote=true,
    language=C,
    classoffset=0, morekeywords={w1, w2, w3, :sem}, keywordstyle=\color{black},
    classoffset=1, morekeywords={add, X, _R, LSL}, keywordstyle=\color{purple},
    classoffset=2, morekeywords={__array}, keywordstyle=\color{codegreen},
    classoffset=0,
}
\DeclareSIUnit{\pp}{\%pt}

\newcommand{\BNFALT}{\;\;|\;\;}
\newcommand{\BNFALTL}{|\;\;}
\newcommand{\bvtype}[1]{\ensuremath{\textsf{bv}\langle{#1}\rangle}}
\newcommand{\decl}[2]{\ensuremath{{#1}: {#2}}}
\newcommand{\assrt}[1]{\ensuremath{\textsf{assert}~{#1}}}
\newcommand{\assume}[1]{\ensuremath{\textsf{assume}~{#1}}}
\newcommand{\goto}[1]{\ensuremath{\textsf{goto}~{#1}}}
\newcommand{\lab}[1]{\ensuremath{\textsf{label}~{#1}}}
\newcommand{\havoc}[1]{\ensuremath{\textsf{havoc}~{#1}}}
\newcommand{\bvop}[1]{\ensuremath{\oplus({#1})}}

\newcommand{\ultimate}{UAutomizer\xspace}
\newcommand{\Ozero}{\texttt{-O0}\xspace}
\newcommand{\Otwo}{\texttt{-O2}\xspace}
\newcommand{\basil}{Basil\xspace}
\newcommand{\code}[1]{{\ttfamily\footnotesize #1}}
\newcommand{\sat}{SAT\xspace}
\newcommand{\unsat}{UNSAT\xspace}
\newcommand{\trto}{\ensuremath{\rightsquigarrow}}
\newcommand{\trbv}[2]{\ensuremath{\llbracket{#1}\rrbracket{#2}}}

\newcommand{\hdr}[2]{%
\flushleft
\textbf{#1} \hfill {#2} \\
\centering
}

\newcommandx{\sidebyside}[5][1=.45,2=.45,3=b]{%
\begin{minipage}[#3]{#1\linewidth}
\centering
{#4}
\end{minipage}\hfill
\begin{minipage}[#3]{#2\linewidth}
\centering
{#5}
\end{minipage}%
}

\newcommandx{\threesidebyside}[7][1=.3,2=.3,3=.3,4=b]{%
\begin{minipage}[#4]{#1\linewidth}
\centering
{#5}
\end{minipage}\hfill
\begin{minipage}[#4]{#2\linewidth}
\centering
{#6}
\end{minipage}\hfill
\begin{minipage}[#4]{#3\linewidth}
\centering
{#7}
\end{minipage}%
}

\newcommandx{\foursidebyside}[8][1=.225,2=.225,3=.225,4=.225]{%
\begin{minipage}[b]{#1\linewidth}
\centering
{#5}
\end{minipage}\hfill
\begin{minipage}[b]{#2\linewidth}
\centering
{#6}
\end{minipage}\hfill
\begin{minipage}[b]{#3\linewidth}
\centering
{#7}
\end{minipage}\hfill
\begin{minipage}[b]{#4\linewidth}
\centering
  {#8}
\end{minipage}%
}

\begin{document}
\title{Bit-Vector CHC Solving for Binary Analysis and Binary Analysis for Bit-Vector CHC Solving}
\titlerunning{Bit-Vector CHC Solving for Binary Analysis and Vice Versa}

\ifmyanon\else
\author{Aaron Bembenek\orcidID{0000-0002-3677-701X} \and
Toby Murray\orcidID{0000-0002-8271-0289}}
\authorrunning{A. Bembenek and T. Murray}
\institute{School of Computing and Information Systems\\
The University of Melbourne, Parkville VIC 3010, Australia\\
\email{\{aaron.bembenek,toby.murray\}@unimelb.edu.au}}
\fi

\maketitle              %
\begin{abstract}

For high-assurance software, source-level reasoning is insufficient: we need binary-level guarantees.
Despite constrained Horn clause (CHC) solving being one of the most popular forms of automated verification, prior work has not evaluated the viability of CHC solving for binary analysis.
To fill this gap, we assemble a pipeline that encodes binary analysis problems as CHCs in the SMT logic of quantifier-free bit vectors, and show that off-the-shelf CHC solvers achieve reasonable success on binaries compiled from 983 C invariant inference benchmarks: a portfolio solves \SI{59.5}{\percent} and \SI{66.0}{\percent} of the problems derived from the unoptimized and optimized binaries, respectively---roughly equal to the success rate of a leading C verifier on the source code (\SI{60.1}{\percent}).
Moreover, we show that binary analysis provides a valuable source of bit-vector CHC benchmarks (which are in short supply): binary-derived problems differ from existing benchmarks both structurally and in solver success rates and rankings.
Augmenting CHC solving competitions with binary-derived benchmarks
will encourage solver developers to improve bit-vector reasoning, in turn making CHC solving a more effective tool for binary analysis.

\end{abstract}

\section{Introduction}

This paper applies bit-vector constrained Horn clause (CHC) solving to binary analysis, with two goals: 1) to assess the viability of CHC solving as a tool for binary analysis, and 2) to use binaries as a source for deriving interesting bit-vector CHC problems for CHC solver competitions.
These goals support each other: doing CHC-based binary analysis leads to binary-derived bit-vector benchmarks that can be added to CHC solver competitions; in turn, solver developers are incentivized to improve bit-vector reasoning in ways that make CHC-based binary analysis more feasible, leading to more binary-derived benchmarks.

Binary analysis is a critical application domain for program reasoning and verification.
Even if source code is available, we might need strong guarantees about the code that will actually run---i.e., the binary.
Important properties that hold at the source level might not hold at the binary level, due to a buggy compiler, compromised build infrastructure (e.g., the SolarWinds software supply chain attack),
or the correctness-security gap in compiler optimizations~\cite{bal10,DSilva2015Correctness}.

The fact that binary analysis is difficult to do by hand, and that there are not enough human experts to do it, motivates the development of automated techniques to aid human analysts.
From the perspective of minimizing human effort, automating binary analysis is arguably more crucial than automating source analysis; e.g., whereas it is conceivable that a programmer could manually figure out a source-level loop invariant, doing so in low-level binary code would be substantially more burdensome.
However, it is a challenging domain for automated reasoning tools, as binaries lack high-level abstractions like program variables, types, data structures, and structured control flow, and are full of low-level bit-vector operations, which are expensive to reason about precisely.

This last point---the prevalence of low-level bit-vector operations---is perhaps the reason why CHC solving, despite being one of the most popular forms of automated verification, has not been applied to binary analysis:
although CHCs have become a de facto standard intermediate representation for model checking, historically there has not been much focus in the CHC space on bit-vector reasoning, and few CHC solvers support the SMT theory of bit vectors~\cite{Hojjat2018ELDARICA}.
However, the landscape is changing, as CHC-COMP 2025 introduced a bit-vector reasoning track.
Our work is intended, in part, to encourage the development both of this competition track, and bit-vector reasoning in CHC solvers more broadly.

As our primary contributions, we present three findings:

\medskip
\noindent\textbf{1) Binary Safety Problems Can Be Encoded as Bit-Vector CHCs}~
We assemble a pipeline that extracts bit-vector CHC problems from AArch64 Linux binaries.
We use the pipeline to generate 1966 bit-vector CHC problems from the \Ozero and \Otwo
binaries of 983 C benchmarks for safety invariant inference.

\medskip
\noindent\textbf{2) CHC Solving Is Reasonably Effective for Binary Analysis}~
We evaluate three off-the-shelf CHC solvers~\cite{Komuravelli2014SMT,Hojjat2018ELDARICA,Toth2017Theta} on the 1966 binary-derived problems, and find that a portfolio of the solvers succeeds \SI{62.8}{\percent} of the time---performance comparable to that of a leading C verifier~\cite{Heizmann2013Software} on the original source code.
Our experiments demonstrate the utility of binary analysis compared to just source-level reasoning: we find cases where the source program is unsafe (because of undefined behavior), but one or more binary is safe. 

\medskip
\noindent\textbf{3) Binaries Provide Diverse Bit-Vector CHC Benchmarks}~
We show that binaries are a good source of interesting bit-vector CHC competition problems: binary-derived problems differ meaningfully, both from existing CHC-COMP benchmarks, and between different categories of binaries (e.g., unoptimized vs optimized).
A classifier can use structural features to distinguish between CHC-COMP problems and four categories of binary-derived problems with \SI{91}{\percent} accuracy, and CHC solver performance varies among these different problem classes.

\medskip
\noindent\textbf{Implications}~
Our work finds that binary-derived CHC problems are structurally different from existing bit-vector CHC benchmarks, but off-the-shelf solvers still perform reasonably well on binary-derived problems.
Thus, it might be possible to improve the performance of CHC solving on binary analysis by developing solvers with a mind towards features that distinguish binary-derived problems.
We also find that problem structure and solver performance vary among different categories of binary-derived benchmarks (e.g., unoptimized vs optimized binaries).
This confirms that the effectiveness of CHC solving for binary analysis is ultimately a function of not only the source code behind a binary, but also the compilation strategy.
Moreover, we can generate different flavors of binary-derived bit-vector CHC problems for competitions by varying both the source code and the compiler options that are used to create the binary.

\medskip
\noindent\textbf{Roadmap}~
Section~\ref{sec:background} gives background on CHC solving.
Section~\ref{sec:pipeline} describes our pipeline for extracting CHC problems from binaries and the resulting benchmark suite.
Section~\ref{sec:eval} evaluates CHC solver performance on binary analysis.
Section~\ref{sec:eval_benchmarks} demonstrates the structural and performance diversity of binary-derived CHC problems.
Section~\ref{sec:related} covers related work, and Section~\ref{sec:limitations} discusses limitations.

\medskip
\noindent\textbf{Data Availability}~
Our code, experiments, and results are publicly available~\cite{benchmarks}.

\section{Background: CHC Solving}\label{sec:background}

Over the last decade, constrained Horn clauses (CHCs) have become a popular representation for verification problems~\cite{Bjoerner2012Program,Bjoerner2015Horn,Gurfinkel2022Program}, and many program analysis tools now accept or produce verification conditions in the form of CHCs---indeed, there are CHC-based verifiers for C~\cite{Ernst2023Korn}, Java~\cite{Kahsai2016JayHorn}, Rust~\cite{Matsushita2021RustHorn}, Solidity~\cite{Alt2022SolCMC}, and LLVM IR~\cite{Gurfinkel2015SeaHorn}.
A CHC is a first-order logic formula of the form
$$\forall X.~\big(p_1(\mathbf{x_1}) \wedge \cdots \wedge ~p_n(\mathbf{x_n}) \wedge \phi \Longrightarrow h\big)$$
where $X$ is the set of all variables appearing in the implication, each $p_i(\mathbf{x_i})$ is an uninterpreted predicate $p_i$ applied to a sequence of variables $\mathbf{x_i}$ (here {\bf bold} indicates repetition), $\phi$ is an interpreted formula from a particular SMT logic (e.g., quantifier-free linear arithmetic, \code{QF\_LIA}), and the head $h$ is either an uninterpreted predicate $p(\mathbf{x})$ or the symbol $\bot$ (denoting false).
Following terminology from logic programming, a clause where the head is $\bot$ is known as a ``query''.

A solution to a conjunction of CHCs is a map from uninterpreted predicates to interpreted formulas that makes each CHC in the system valid.
Typically, the formulas in the solution must also fall within a particular SMT logic.
The formulas for the predicates act as safety invariants; hence, a \sat system of CHCs represents a safe program, and an \unsat system an unsafe program.
In this way, queries act as assertions: the premise of a query must always be false.

Techniques for solving CHCs range from counterexample-guided abstraction refinement (CEGAR)~\cite{Grebenshchikov2012Synthesizing,Hojjat2018ELDARICA,Blicha2023Golem} to IC3/PDR~\cite{Hoder2012Generalized,Komuravelli2013Automatic,Blicha2023Golem} to data-driven learning~\cite{Ezudheen2018Horn,Fedyukovich2018Solving,Zhu2018Data}.
Despite the wide range of solving techniques, and the importance of bit-vector reasoning for software verification, few solvers support bit-vector reasoning (i.e., where the logic of interpreted formulas is \code{QF\_BV})~\cite{Hojjat2018ELDARICA}.
CHC solvers that do support bit vectors include Spacer, Eldarica, and Theta.
Spacer~\cite{Komuravelli2014SMT} implements an extended IC3/PDR algorithm that maintains both over- and under-approximations of solutions.
Eldarica~\cite{Hojjat2018ELDARICA} is based on CEGAR and predicate abstraction; it uses convergence heuristics like acceleration and interpolation abstraction to more quickly reach a solution.
Theta~\cite{Toth2017Theta} is a generic model checking framework that encodes a system of CHCs as a control flow automaton and then checks for error reachability using diverse model checking algorithms (e.g., bounded model checking, k-induction, interpolation-based techniques).  

\section{From Binaries to Bit-Vector CHC Problems}\label{sec:pipeline}

We overview our pipeline for generating bit-vector CHC problems from binaries (Section~\ref{sec:gen_pipeline}), give a technical description of the CHC encoding we use (Section~\ref{sec:encoding}), and describe the CHC benchmarks we generated
(Section~\ref{sec:seahorn_benchmarks}).

\subsection{Overview}\label{sec:gen_pipeline}

\begin{figure}[!b]
\centering
\begin{tikzpicture}[
    node distance=1.5cm and 2.2cm,
    box/.style={
        rectangle,
        rounded corners=2pt,
        minimum width=1.5cm,
        minimum height=1.2cm,
        text centered,
        draw=black!60,
        fill=boxcolor,
        font=\small,
        text width=1.2cm,
        align=center
    },
    ourbox/.style={
        rectangle,
        rounded corners=2pt,
        minimum width=1.5cm,
        minimum height=1.2cm,
        text centered,
        draw=black!60,
        fill=orangebox,
        font=\small,
        text width=1.2cm,
        align=center
    },
    arrow/.style={
        -Stealth,
        thick,
        color=arrowcolor
    },
    label/.style={
        font=\small\itshape,
        text=textcolor,
        align=center
    }
]

\node[box] (lifting) {Lift};

\node[ourbox, right=of lifting] (encoding) {Encode};

\node[box, right=of encoding] (simplifying) {Simplify};

\draw[arrow] ($(lifting.west) + (-1.5,0)$) -- (lifting.west) 
    node[label, pos=0.5, above] {Binary};

\draw[arrow] (lifting.east) -- (encoding.west)
    node[label, pos=0.5, above, text width=2.2cm] {Lifted binary\\(Boogie)};

\draw[arrow] (encoding.east) -- (simplifying.west)
    node[label, pos=0.5, above, text width=1.8cm] {Initial\\CHCs};

\draw[arrow] (simplifying.east) -- ($(simplifying.east) + (1.5,0)$)
    node[label, pos=0.5, above, text width=1.5cm] {Final\\CHCs};

\end{tikzpicture}
\caption{We assembled a pipeline for deriving bit-vector CHC problems from binaries: the binary is lifted to a Boogie program, the Boogie program is encoded as CHCs, and then the CHCs are simplified. The first and third stages use off-the-shelf components (blue); the second stage is our own code (orange).}
\label{fig:pipeline}
\end{figure}
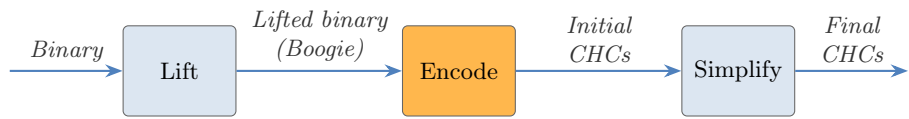

The benchmark generation pipeline consists of three main stages: binary lifting, encoding the lifted binary into CHCs, and simplifying the CHCs (Figure~\ref{fig:pipeline}).
We describe each stage in turn.
The first and third stages use off-the-shelf components; the second stage---encoding the lifted binary as CHCs---is our own code.

\subsubsection{Binary Lifting}

The pipeline begins with lifting an AArch64 Linux binary into an intermediate representation that is easy to encode as CHCs.
This happens in three steps.
First, the binary is disassembled via DDisasm~\cite{FloresMontoya2020Datalog} into the GTIRB format~\cite{Schulte2019GTIRB}.
Second, the gtirb-semantics tool~\cite{gtirbsemantics2025} enriches the GTIRB code with semantics derived from the ARMv8 formal specification via ASLp~\cite{Lam2023Lift}.
Third, the \basil binary analysis platform~\cite{basil2025} analyzes the enriched GTIRB code (along with the output of the readelf utility run on the binary).
In addition to performing generic simplifications like constant propagation, \basil attempts to recover higher-level abstractions.
For example, \basil translates operations over status flags (e.g., the carry flag \code{CF}) into bit-vector comparisons, and uses memory region inference~\cite{bay26} to turn memory accesses into reads and writes of bit-vector variables.
Memory accesses that cannot be resolved to operations over bit-vector variables are left as array operations over a flat representation of memory.
As our focus is on generating (array-free) bit-vector CHC benchmarks, our pipeline rejects any binary where \basil is unable to simplify away all memory accesses.

\subsubsection{Encoding into CHCs}

\basil outputs the simplified lifted binary as a Boogie program; Boogie is a widely used, statically typed intermediate verification language designed for SMT-based verification~\cite{Barnett2005Boogie}.
In the second phase of the pipeline, we encode this Boogie program as a system of CHCs (Section~\ref{sec:encoding}).

\subsubsection{Simplification}

In the final phase, we simplify each benchmark using the formatting script from CHC-COMP.%
\footnote{\url{https://github.com/chc-comp/scripts/tree/master/format}}
This rewrites the benchmark into a standard form and simplifies it using Z3's~\cite{Moura2008Z3} simplification tactics.
Simplification can reveal that a benchmark is trivially satisfiable; if so, we drop that benchmark.

\subsection{CHC Encoding}\label{sec:encoding}

\begin{figure}[!b]
    \centering
    \sidebyside[0.45][0.40][t]{
        \[\begin{array}{@{}l@{}rcl@{}}
            \text{Programs} & prog & ::= & P^+ \\
            \text{Procedures} & P & ::= & f(d^*)~\mathsf{returns}~(d^*)~\{ d^* B^+ l \} \\
            \text{Declarations} & d & ::= & \decl{x}{\bvtype{k}} \\
            \text{Block} & B & ::= & l: s^*~;~\goto{l^+} \\
            \text{Statements} & s & ::= & \assrt{e} \BNFALT \assume{e} \\
            & & & \BNFALTL x := e \BNFALT \havoc{x} \\
            & & & \BNFALTL x^* := \mathsf{call}~f(e^*)\\
            \text{Expressions} &~e & ::= & x \BNFALT v \BNFALT \bvop{e^+}
        \end{array}\]
    }
    {
        \[\begin{array}{@{}l@{}rcl@{}}
        \multicolumn{4}{l}{\textbf{Namespaces}}                  \\ \hline
            \text{Program labels}    & l        & \in & \mathrm{Label} \\
            \text{Procedure names}     &~f        & \in & \mathrm{PVar} \\
            \text{Bit-vector variables}     &~~x, y        & \in & \mathrm{BVar}     \\
            \text{Bit-vector literals}      & v        & \in & \mathbb{B}^+ \\
            \text{Positive integers} & k & > & 0 
        \end{array}\]
    }
    \caption{The lifted binaries we encode as CHCs are Boogie programs described by this (stylized) grammar.
    Boolean values are treated as bit vectors of size one.}
    \label{fig:boogie_grammar}
  \end{figure}

\basil outputs the lifted binary as a Boogie program~\cite{Barnett2005Boogie} consisting of a set of procedures (Figure~\ref{fig:boogie_grammar}).%
\footnote{Technically, the Boogie program that \basil generates might also contain global variables. In an initial pass, our code rewrites the program to remove global variables.}
A procedure declares both in- and out-parameters.
Each procedure consists of local bit-vector variables, a sequence of blocks, and a terminal label (the procedure exit).
A block is a label, a sequence of statements, and then a goto to one or more labels in the procedure; gotos with multiple targets indicate non-deterministic control flow.
A statement is an assertion, an assumption, an assignment, a procedure call, or a variable havoc (non-deterministic assignment).
A conditional is encoded by jumping non-deterministically to two blocks: one assumes the condition, the other assumes its negation.
A bit-vector expression is a variable, literal, or operation.
A variable represents either a machine register or a higher-level variable inferred by \basil's memory region analysis.

\begin{figure}[!t]

\hdr{Additional Namespaces}{}
\vspace{5pt}

SMT conjunctions $C$, sets of Horn clauses $\mathcal{H}$, variable maps $\Delta \in \mathrm{BVar} \rightarrow \mathrm{BVar}$

\hdr{Expression Encoding}{\fbox{$\trbv{e}{\Delta} = e$}}

\medskip

\threesidebyside[0.2][0.2][0.55]{
\infrule{}{\trbv{v}{\Delta} = v}
}{
\infrule{y = \Delta(x)}{\trbv{x}{\Delta} = y}
}{
\infrule{
    \forall i \in [1, n].~\trbv{e_i}{\Delta} = {e_i}'
}{
    \trbv{\oplus(e_1, \dots, e_n)}{\Delta} = \oplus({e_1}', \dots, {e_n}')
}
}

\hdr{Statement Encoding}{\fbox{$\Delta, C \vdash s \trto \mathcal{H} : \Delta, C$}}

\medskip

\sidebyside{
\infrule[Assume]{
    \trbv{e}{\Delta} = e'
}{
    \Delta, C \vdash \assume{e} \trto \emptyset : \Delta, C \wedge e'
}
}{
\infrule[Assert]{
    \trbv{e}{\Delta} = e'
    \quad
    \mathcal{H} = \{ C \wedge \neg e' \implies \bot \}
}{
    \Delta, C \vdash \assrt{e} \trto \mathcal{H} : \Delta, C \wedge e'
}
}

\bigskip

\sidebyside[0.4][0.55]{
\infrule[Havoc]{
    y = fresh(x) \quad \Delta' = \Delta[x \mapsto y]
}{
    \Delta, C \vdash \havoc{x} \trto \emptyset : \Delta', C
}
}{
\infrule[Assign]{
    y = fresh(x) \quad \Delta' = \Delta[x \mapsto y] \quad \trbv{e}{\Delta} = e'
}{
    \Delta, C \vdash x := e \trto \emptyset : \Delta', C \wedge y = e'
}
}

\infrule[Call]{
    \forall i \in [1, n].~\trbv{e_i}{\Delta} = {e_i}'
    \quad
    \forall j \in [1, m].~y_j = fresh(x_j)
    \\ 
    \Delta' = \Delta[x_1 \mapsto y_1, \dots, x_m \mapsto y_m]
    \\
    \mathcal{H} = \{ C \implies \mathsf{enter}\langle f \rangle({e_1}', \dots, {e_n}') \}
    \quad
    C' = C \wedge \mathsf{exit}\langle f \rangle({e_1}', \dots, {e_n}', y_1, \dots, y_m)
}{
    \Delta, C \vdash x_1, \dots, x_m := \mathsf{call}~f(e_1, \dots, e_n) \trto \mathcal{H} : \Delta', C'
}

\hdr{Block Encoding}{\fbox{$\vdash B \trto \mathcal{H}$}}

\infrule[Block]{
    \Delta_0 = \lambda x.x \quad
    C_0 = pred(l, \Delta_0) \quad
    \forall i \in [1, m].~\Delta_{i-1}, C_{i-1} \vdash s_i \trto \mathcal{H}_i : \Delta_i, C_i \\
    \mathcal{H}_{goto} = \{ C_m \implies pred(l_j, \Delta_m) \mid j \in [1, n] \}
}{
    \vdash l: s_1, \dots, s_m~;~\goto{l_1, \dots, l_n} \trto \mathcal{H}_1 \cup \cdots \cup \mathcal{H}_m \cup \mathcal{H}_{goto}
}

\caption{The three judgments given here define how we encode Boogie expressions, statements, and blocks as bit-vector CHCs.}\label{fig:encoding}
\end{figure}

We define the encoding of Boogie expressions, statements, and blocks via three judgments (Figure~\ref{fig:encoding}) that build on the general formula for encoding imperative programs as CHCs~\cite{Bjoerner2015Horn}.
Expressions are encoded using the straightforward judgment $\trbv{bv}{\Delta} = bv$, where the function $\Delta$ maps a variable in the Boogie program to a variable in the SMT expression.
Our fragment of Boogie and our SMT target language share the same language of expressions.

Statements are encoded via the judgment $\Delta, C \vdash s \trto \mathcal{H}: \Delta, C$.
We use the metavariable $C$ to denote a logical conjunction (a clause premise under construction) and the metavariable $\mathcal{H}$ to denote a set of Horn clauses, in which all variables are implicitly bound by an outer universal quantifier.
The judgment $\Delta, C \vdash s \trto \mathcal{H}: \Delta', C'$ intuitively means that, given variable map $\Delta$ and partial clause $C$, encoding the statement $s$ results in clause set $\mathcal{H}$, an updated variable map $\Delta'$, and updated partial clause $C'$ (which may have a new conjunct).

Assumptions augment the partially constructed clause $C$ with the assumed value (\rn{Assume}).
Assertions do the same, but additionally generate a query, stating that the partial clause (conjunction) $C$ implies the asserted value (\rn{Assert}).
The statement \havoc{x} generates a fresh variable $y$ and updates $\Delta$ to map $x$ to $y$ (\rn{Havoc}); we assume a helper $fresh(x)$ that generates a fresh SMT variable for Boogie variable $x$.
Assignments do the same, and also update the partially constructed clause $C$ to constrain $y$ with the assigned value (\rn{Assign}).

Each program label is associated with an SMT predicate.
The function $pred(l, \Delta)$ returns the predicate for label $l$, using $\Delta$ to map its formal parameters to arguments.
The predicate for the first label in a procedure $f$ is named $\mathsf{enter}\langle f \rangle$, and its parameters correspond to the in-parameters to the procedure.
Similarly, the predicate for the last label is named $\mathsf{exit}\langle f \rangle$, and its parameters correspond to the in- and out-parameters to procedure $f$.
For all other predicates, the parameters correspond to the procedure's in-parameters and any program variables that are live at that label.
The in-parameters act as a call context.

Procedure calls are encoded using the predicates for the called procedure's entry and exit (\rn{Call}).
In particular, a Horn clause is generated stating that the procedure entry is reachable with particular arguments, and the partial clause $C$ is extended to include the exit predicate for the called procedure (which captures the values returned from the call, as well as the arguments for the call).

The judgment $\vdash B \trto \mathcal{H}$ defines how a block $B$ is encoded as a set of CHCs $\mathcal{H}$.
Each statement of the block is encoded in order, using an initial partial clause $C_0$ that contains only a predicate for that block's label (\rn{Block}).
To capture control flow from a block, the partial clause $C_m$ resulting from encoding the block's body is used to construct a Horn clause for each target in the block's goto statement; these clauses refer to the predicate for each target label.

A program is encoded by encoding all the blocks in its procedures, and adding a clause stating that the \code{main} procedure is reachable.

\begin{example}\label{ex:encoding}
Consider encoding the following Boogie block as a system of CHCs, where the prefix $\mathsf{bvs}$ indicates the signed version of a bit-vector operation:
\begin{lstlisting}
foo:
  assume bvsge(x, y)
  x := bvadd(x, 001)
  assert bvsgt(x, y) ;
  goto bar, baz
\end{lstlisting}
We assume that the predicate for each label has the same name as the label, and that each predicate has parameters corresponding to the three-bit program variables $\mathsf{x}$ and $\mathsf{y}$.
The block is encoded as the following three CHCs:
\begin{align*}
&\forall \mathsf{x}, \mathsf{y}, \mathsf{x}'.~\mathsf{foo}(\mathsf{x}, \mathsf{y}) \wedge \mathsf{bvsge}(\mathsf{x}, \mathsf{y}) \wedge \mathsf{x}' = \mathsf{bvadd}(\mathsf{x}, 001) \wedge \neg \mathsf{bvsgt}(\mathsf{x}', \mathsf{y}) \implies \bot \\
&\forall \mathsf{x}, \mathsf{y}, \mathsf{x}'.~\mathsf{foo}(\mathsf{x}, \mathsf{y}) \wedge \mathsf{bvsge}(\mathsf{x}, \mathsf{y}) \wedge \mathsf{x}' = \mathsf{bvadd}(\mathsf{x}, 001) \wedge \mathsf{bvsgt}(\mathsf{x}', \mathsf{y}) \implies \mathsf{bar}(\mathsf{x}', \mathsf{y}) \\
&\forall \mathsf{x}, \mathsf{y}, \mathsf{x}'.~\mathsf{foo}(\mathsf{x}, \mathsf{y}) \wedge \mathsf{bvsge}(\mathsf{x}, \mathsf{y}) \wedge \mathsf{x}' = \mathsf{bvadd}(\mathsf{x}, 001) \wedge \mathsf{bvsgt}(\mathsf{x}', \mathsf{y}) \implies \mathsf{baz}(\mathsf{x}', \mathsf{y})
\end{align*}
The first CHC is the query produced for the assertion statement.
The second two CHCs encode the non-deterministic control flow at the end of the block.
The variable $\mathsf{x}'$ is a fresh ``version'' of the variable $\mathsf{x}$, used to hold its new value after the assignment.
Table~\ref{tab:example} gives the values for the $\Delta_k$ and $C_k$ variables used to instantiate the \rn{Block} rule on this example.
If the system of CHCs is satisfiable, the CHC solver will assign a formula in the SMT logic of \code{QF\_BV} to each uninterpreted predicate.
For example, the interpretation $\mathsf{foo}(x, y) \equiv \mathsf{bvslt}(x, 011)$ is sufficient for ensuring that the query does not fail due to integer overflow.
\end{example}

\begin{table}[!t]
\caption{Evolution of variable mapping ($\Delta_k$) and partial clause ($C_k$) when encoding the statements in Example~\ref{ex:encoding} using the \rn{Block} rule}\label{tab:example}
\centering
\begin{tabular}{l@{\hspace{1em}}l@{\hspace{1em}}l}
    \toprule
    $k$ & $\Delta_k$ & $C_k$ \\
    \midrule
    0 & $\lambda z.z$ & $\mathsf{foo}(\mathsf{x}, \mathsf{y})$ \\
    1 & $\lambda z.z$ & $\mathsf{foo}(\mathsf{x}, \mathsf{y}) \wedge \mathsf{bvsge}(\mathsf{x}, \mathsf{y})$ \\
    2 & $\lambda z.(z = \mathsf{x})\;?\;\mathsf{x}':z$ & $\mathsf{foo}(\mathsf{x}, \mathsf{y}) \wedge \mathsf{bvsge}(\mathsf{x}, \mathsf{y}) \wedge \mathsf{x}' = \mathsf{bvadd}(\mathsf{x}, 001)$ \\
    3 & $\lambda z.(z = \mathsf{x})\;?\;\mathsf{x}':z$ & $\mathsf{foo}(\mathsf{x}, \mathsf{y}) \wedge \mathsf{bvsge}(\mathsf{x}, \mathsf{y}) \wedge \mathsf{x}' = \mathsf{bvadd}(\mathsf{x}, 001) \wedge \mathsf{bvsgt}(\mathsf{x}', \mathsf{y})$ \\
    \bottomrule
\end{tabular}
\end{table}

\subsection{Benchmarks}\label{sec:seahorn_benchmarks}

We used this pipeline to generate 1966 binary-derived bit-vector CHC problems, corresponding to paired \Ozero and \Otwo binaries for 983 C programs that range in length from 31 SLOC to 6150 SLOC (median/mean: 846/1033 SLOC).

We started with the 1469 C programs in the SeaHorn-LinearArbitrary benchmark suite~\cite{Zhu2018Data}.
These programs contain loops or recursion, and the goal of a verifier is to show that error locations (such as a \code{\_\_VERIFIER\_error()} call) are unreachable.
The C benchmarks were curated by Zhu et al. from prior work, and fall into two main categories: short, handwritten challenge problems~\cite{Garg2016Learning,Padhi2016Data}, and long SV-COMP benchmarks based on C code generated by, e.g., model checkers.
We standardized the C programs and attempted to make each one compile using gcc (v14.2.1) and the \Ozero and \Otwo flags; however, not all compiled successfully.
\basil itself did not complete on all binaries, and the CHC-COMP formatting script timed out on some problems and declared others trivial.
In order that the suite might be balanced, we kept only problems where the pipeline completed successfully for both the \Ozero and \Otwo binaries.
We ended up with 1966 CHC problems; 1192 of them (\SI{61}{\percent}) are from binaries for SV-COMP benchmarks.

As an alternative to the SeaHorn-LinearArbitrary benchmark suite, we also considered more recent benchmark suites used to evaluate neural-network-based invariant inference.
However, the suites we inspected~\cite{Si2020Code2Inv,Kamath2024Leveraging,Pirzada2024LLM,Wu2024LLM} contain only short C programs---typically less than 50 SLOC, and no more than $\sim\!200$ SLOC.

\section{Evaluating CHC Solvers on Binary Analysis}\label{sec:eval}

We use our set of binary-derived CHC problems to evaluate the potential of CHC solving for binary analysis, answering the following research questions:
\begin{rqlist}
    \item How do different CHC solvers perform on binary-derived problems?
    \item What effect does compiler optimization level have on CHC solver success?
    \item How does the success rate of binary-level reasoning compare to the success rate of source-level reasoning?
\end{rqlist}
Our results---that a portfolio of solvers is able to solve \SI{62.8}{\percent} of the binary-derived CHC problems---suggest that CHC solving is a promising technology for binary analysis.
Indeed, on the problems in our experiments, CHC solving at the binary level has a success rate similar to source-level reasoning (\SI{60.1}{\percent}).

\subsection{Experimental Setup}

We evaluate the three bit-vector CHC solvers described in Section~\ref{sec:background}: Eldarica (commit 92860f1)~\cite{Hojjat2018ELDARICA}, Spacer (Z3 v4.15.4)~\cite{Komuravelli2014SMT}, and Theta (v6.21.4)~\cite{Toth2017Theta}.
We chose Eldarica and Theta because of their first- and second-place finishes, respectively, in the bit-vector category of CHC-COMP 2025,
and Spacer because it is the default CHC solver in the popular SMT solver Z3~\cite{Moura2008Z3}.
We use Eldarica and Theta in their CHC-COMP 2025 configurations, and Spacer in its default configuration.
For source-level reasoning, we use Ultimate Automizer (v0.3.0)~\cite{Heizmann2013Software}.
An ``Unknown'' result indicates that a tool crashed or timed out on a benchmark.

We ran all experiments on a Red Hat Enterprise Linux 9.6 cluster where each experiment was allocated 8 Intel(R) Xeon(R) Gold 6448H CPU cores and 30GB of memory.
For Eldarica, Theta, and \ultimate, we set a max JVM heap size of 24GB, reserving the remaining memory for external processes.
We used a time limit of 15 minutes for each experiment.
We determined this limit based on the results of CHC-COMP 2025: of the 440 cases where Eldarica or Theta succeeded in the bit-vector category, 428 cases (97\%) completed in under 10 minutes.

\subsection{Solver Comparison}

\begin{figure}[!b]
\centering
\includegraphics[width=\textwidth]{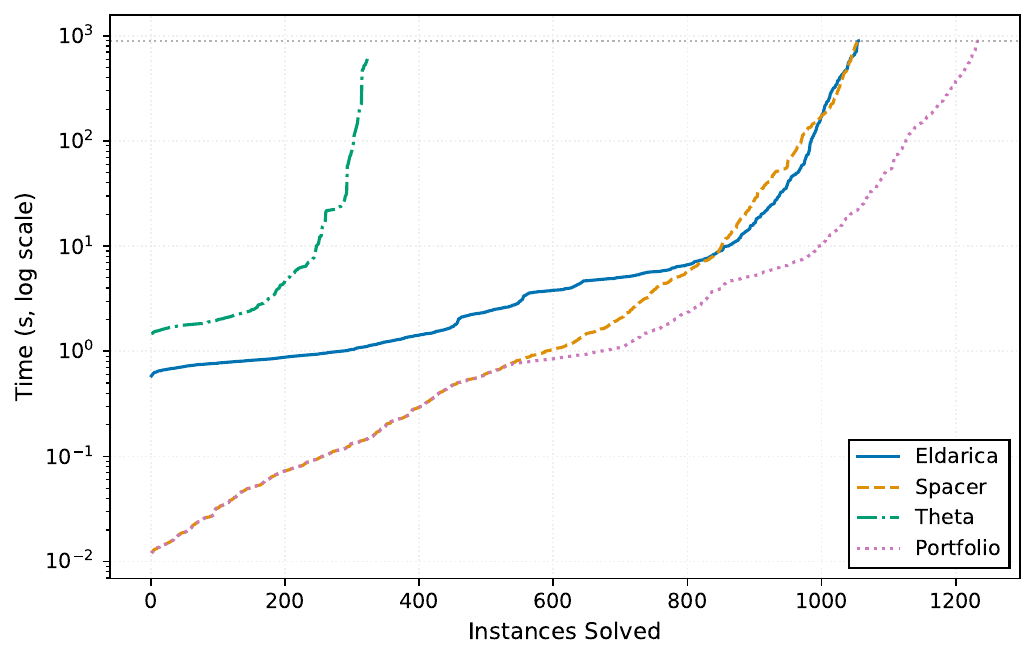}
\caption{Eldarica and Spacer solve almost the same number of benchmarks, while a portfolio of the two solvers solves \SI{9}{\pp} more than either individual solver.}
\label{fig:solver-comparison}
\end{figure}

Eldarica and Spacer achieve nearly identical overall success on the benchmark suite, with 1055 and 1053 instances solved, respectively (\figurename~\ref{fig:solver-comparison}).
Of these, 874 instances are solved in common.
Theta is less successful overall, solving only 323 instances; 171 of these instances are also solved by Spacer and Eldarica.
Spacer's median wall-clock time on solved instances is about $3.5\times$ faster than the median times for Eldarica and Theta (\SI{0.72}{\second} vs \SI{2.62}{\second} and \SI{2.77}{\second}, respectively).

While many solved benchmarks are solved by two or more solvers,
352 benchmarks are solved by only one solver: 155 by Eldarica, 125 by Spacer, and 72 by Theta.
This indicates that the solvers are complementary on these binary-derived bit-vector CHC problems.
A virtual portfolio of Spacer and Eldarica solves 1234 (\SI{62.8}{\percent}) of the benchmarks with a median wall-clock time of \SI{0.88}{\second} on solved instances.
In the remainder of this section, we assume this portfolio is used.
We omit Theta from the portfolio as we found that on 33 benchmarks it gives results that are inconsistent with Spacer, Eldarica, or both (in 4 cases).
Given that Spacer and Eldarica never disagree, and that a previous version of Theta gave incorrect results on 27 instances during CHC-COMP 2025,%
\footnote{\url{https://github.com/chc-comp/chc-comp25-benchmarks/pull/7}}
we opt to trust Spacer and Eldarica over Theta.
That being said, the ground truth is not known for each benchmark, and our assumption could be incorrect in some cases.

\begin{tcolorbox}[rqanswerbox]
    \textbf{RQ1 Answer:} Eldarica solves \SI{53.7}{\percent}, Spacer \SI{53.6}{\percent}, and Theta \SI{16.4}{\percent} of the 1966 binary-derived bit-vector CHC problems.
    The solvers are complementary: a virtual portfolio combining Eldarica and Spacer solves \SI{62.8}{\percent} of the problems, an improvement of \SI{9}{\pp} over each solver by itself.
\end{tcolorbox}

\subsection{Effect of Compiler Optimization Level}

\begin{table}[!b]
\centering
\caption{Contingency table: \Ozero vs \Otwo results}
\label{tab:optimization-contingency}
\begin{tabular}{l@{\hspace{1em}}rrr@{\hspace{1em}}r}
\toprule
& \multicolumn{3}{c}{\textbf{-O2 Binary}} \\
\textbf{-O0 Binary} & \textbf{SAT} & \textbf{UNSAT} & \textbf{Unknown} & \textbf{Total} \\
\midrule
\textbf{SAT} & 333 & 1 & 43 & 377 \\
\textbf{UNSAT} & 2 & 196 & 10 & 208 \\
\textbf{Unknown} & 49 & 68 & 281 & 398 \\
\midrule
\textbf{Total} & 384 & 265 & 334 & 983 \\
\bottomrule
\end{tabular}
\end{table}

As described in Section~\ref{sec:seahorn_benchmarks}, our suite of binary-derived CHC problems consists of 983 pairs, where each pair corresponds to the \Ozero (unoptimized) and \Otwo (optimized) binaries compiled from a shared source program.
This pairing of binaries allows us to evaluate the effect of optimization level on the effectiveness of CHC solving for binary analysis.
Optimization level \Otwo is a popular choice that enables most optimizations
that do not involve a space-time tradeoff, such as constant propagation and dead
code elimination.
In Section~\ref{sec:struct_diversity}, we show that CHC problems derived from \Ozero and \Otwo binaries differ structurally.

Overall, CHC solving is more effective on \Otwo problems than \Ozero problems (Table~\ref{tab:optimization-contingency}), primarily because more SV-COMP problems are solved at the \Otwo level.
The virtual portfolio solver (combining Spacer and Eldarica) solves 585 \Ozero problems (\SI{59.5}{\percent}) and 649 \Otwo problems (\SI{66.0}{\percent}).
It succeeds on roughly the same number of \sat problems at the \Ozero and \Otwo levels (377 vs 384), but \SI{27}{\percent} more \unsat problems at the \Otwo level (208 vs 265).

Moreover, despite the overall improvement in performance on \Otwo problems, there are still 53 cases where the \Ozero problem is solved and the \Otwo one is not (43 \sat instances and 10 \unsat instances).
There are also three cases where the \Ozero and \Otwo problems have different results: one where the \Ozero problem is \sat and the \Otwo problem is \unsat, and two vice versa.
In all cases, the source program contains undefined behavior (signed integer overflow or underflow), and the difference between the \Ozero and \Otwo results lies in the compiler taking advantage of this undefined behavior when applying \Otwo optimizations.
This illustrates how the safety of the binary ultimately relies on the choices made by the compiler---a good argument for the importance of binary analysis.

\begin{tcolorbox}[rqanswerbox]
    \textbf{RQ2 Answer:} The CHC solvers are slightly more effective at solving problems derived from optimized binaries, as indicated by a \SI{6.5}{\pp} improvement in solving \Otwo problems compared to \Ozero problems (\SI{66.0}{\percent} vs \SI{59.5}{\percent}).
\end{tcolorbox}

\subsection{Comparison to Source-Level Reasoning}\label{sec:source_comparison}

To compare the success rates of source-level and binary-level reasoning, we ran the Ultimate Automizer verifier (\ultimate)~\cite{Heizmann2013Software} on the 983 C programs that are the source of the binaries we analyze.
\ultimate uses counterexample-guided trace abstraction; we chose it as a representative source-level verifier primarily because of its strong performance in the SV-COMP 2025 verification condition, where it won the integer overflow track and was the overall competition champion~\cite{Beyer2025Improvements}.
\ultimate is also the backend engine used by the CHC solver Unihorn (which does not accept bit-vector CHC problems)~\cite{DeAngelis2023CHC}.
We configure \ultimate to verify not only the unreachability of errors, but also the absence of signed integer overflow and underflow; without this second check, \ultimate can declare that C programs with undefined behavior are safe.

Perhaps surprisingly, binary-level reasoning is as successful as source-level reasoning on the problems we evaluated (albeit we used a portfolio solver for the binaries, and a single solver for the source; Table~\ref{tab:source-binary-contingency-compact}).
\ultimate solved 591 source benchmarks (\SI{60.1}{\percent})---roughly equal to the number of benchmarks solved at the \Ozero level (\SI{59.5}{\percent}), and slightly lower than the number of benchmarks solved at the \Otwo level (\SI{66.0}{\percent}).
Nearly all problems solved at the source level are solved at the \Ozero or \Otwo level (\SI{88.2}{\percent} and \SI{89.0}{\percent} of problems, respectively).

Source-level reasoning has additional proof obligations compared to binary-level reasoning, as \ultimate is required to ensure not only that error locations are unreachable, but also that there is no signed integer overflow or underflow (which is undefined behavior in C).
In some cases, these additional obligations might make verification more difficult; in other cases, they make verification easier, as many benchmarks contain signed integer overflow that is easy to identify, leading to an \unsat result.
Overall, 136 of 281 source-level \unsat results (\SI{48.4}{\percent}) are due to signed integer overflow or underflow.
On the other hand, every \unsat binary-level result is because an error location is proven to be reachable.
Indeed, there are 28 and 30 cases, respectively, where the source program is unsafe because of signed integer overflow/underflow, but the \Ozero or \Otwo problem is \sat.
In these cases, the error location is provably unreachable in the relevant binary, even though the source program contains undefined behavior.

\begin{table}[!t]
\centering
\caption{Contingency tables: source-level vs binary-level results}
\label{tab:source-binary-contingency-compact}
\begin{tabular}{lc@{\hspace{1em}}rrr@{\hspace{1em}}rrr@{\hspace{1em}}c}
\toprule
& & \multicolumn{3}{c}{\textbf{\Ozero Binary}} & \multicolumn{3}{c}{\textbf{\Otwo Binary}} & \\
\cmidrule(r{0.8em}){3-5} \cmidrule(r{0.8em}){6-8}
& & \textbf{SAT} & \textbf{UNSAT} & \textbf{Unknown} & \textbf{SAT} & \textbf{UNSAT} & \textbf{Unknown} & \textbf{Total} \\
\midrule
\multirow{3}{*}{\rotatebox{90}{\textbf{Source}}} & \textbf{SAT} & 289 & 0 & 21 & 285 & 0 & 25 & 310 \\
& \textbf{UNSAT} & 28 & 204 & 49 & 30 & 211 & 40 & 281 \\
& \textbf{Unknown} & 60 & 4 & 328 & 69 & 54 & 269 & 392 \\
\midrule
& \textbf{Total} & 377 & 208 & 398 & 384 & 265 & 334 & 983 \\
\bottomrule
\end{tabular}
\end{table}

\begin{tcolorbox}[rqanswerbox]
    \textbf{RQ3 Answer:} On these benchmarks, source-level reasoning and binary-level reasoning achieve comparable success rates: the C verifier \ultimate succeeds \SI{60.1}{\percent} of the time on the source programs, compared to the portfolio CHC solver succeeding \SI{62.8}{\percent} of the time on the binary-derived problems.
\end{tcolorbox}

\section{Evaluating Binary-Derived CHC Problems}\label{sec:eval_benchmarks}

We assess whether binary-derived benchmarks contribute meaningfully to existing bit-vector CHC benchmark suites---in particular, the bit-vector track of CHC-COMP 2025, which consists of 625 problems, none based on binaries (to our knowledge).
We answer the following research questions in the affirmative:
\begin{rqlist}[start=4]
    \item Do binary-derived CHC problems differ structurally from CHC-COMP benchmarks, and do different binary types produce distinct problems?
    \item Does solver performance vary between CHC-COMP and binary-derived benchmarks, and across different categories of binary-derived problems?
\end{rqlist}

\subsection{Structural Diversity}\label{sec:struct_diversity}

To evaluate structural diversity, we engineer features that characterize the structure of bit-vector CHC problems, and then use these features to train a classifier to successfully distinguish between categories of bit-vector CHC problems.

\subsubsection{Feature Engineering}\label{sec:features}

\begin{table}[!t]
    \caption{29 static features we use to characterize bit-vector CHC problems}\label{tab:features}
    \centering
    \begin{tabular}{p{0.35\linewidth} p{0.6\linewidth}}
        \toprule
        \textbf{Features} & \textbf{Description} \\
        \midrule
        \code{num\_clauses}, \code{num\_linear\_clauses}, \code{num\_nonlin\_clauses}, \code{num\_queries}, \code{density\_rec\_clauses} & The number of clauses, linearly recursive clauses (clauses with one recursive predicate in the premise), non-linearly recursive clauses (clauses with more than one recursive premise predicate), queries, and the fraction of clauses that are recursive \\
        \midrule
        \code{num\_preds}, \code{pred\_arity\_$X$} & The number of predicates, and predicate arity aggregate statistics $X \in \{\texttt{mean}, \texttt{median}, \texttt{max}\}$ \\
        \midrule
        \code{num\_sccs}, \code{scc\_size\_$X$} & The number of strongly connected components (SCCs) in the predicate dependence graph, and SCC size aggregate statistics $X \in \{\texttt{mean}, \texttt{median}, \texttt{max}\}$ \\
        \midrule
        \code{num\_ops}, \code{num\_ops\_per\_clause}, \code{num\_$X$\_ops}, \code{density\_$X$\_ops} & The total number of bit-vector operations, the count averaged across the number of clauses, and the total count and density (as a fraction of total operations) of bit-vector operations in categories $X \in \{ \texttt{struct}, \texttt{bitwise}, \texttt{shift}, \texttt{linear}, \texttt{mul}, \texttt{divmod}, \texttt{comp} \}$ \\
        \bottomrule
    \end{tabular}
\end{table}

We identify 29 features that can characterize bit-vector CHC problems (Table~\ref{tab:features}).
These features are static, and can be extracted from the syntax of a given CHC problem.
The features are split into four groups.

The first group captures information about the number of clauses and the shapes of the clauses, such as the fraction of clauses that are recursive.
A clause is recursive if it has a predicate in its premise that is (perhaps indirectly) defined by the predicate in the conclusion of the clause.
Intuitively, CHCs with more recursive clauses should be more difficult to solve, especially if those clauses are non-linearly recursive (i.e., if they have more than one recursive predicate).

The second group of features captures information about the arity of predicates.
The higher the arity of a predicate, the larger the candidate solution space for the symbolic interpretation of that predicate; thus, CHC problems with higher arity predicates should be harder to solve in general.

The third group captures information about the recursive dependencies between predicates via statistics about the strongly connected components (SCCs) in the predicate dependence graph.
In this graph, nodes represent predicates, and there is an edge from a node for predicate $p$ to the node for predicate $q$ iff there is a clause with predicate $p$ in the premise and predicate $q$ in the conclusion.
An SCC in the predicate dependence graph corresponds to a group of mutually recursive predicates; large SCCs indicate more complex recursive dependencies.

The fourth group captures which bit-vector operations occur in the CHCs.
Here, we partition operations into seven different categories, and include both the absolute count of operations in each category and the fraction of all operations that falls into each category.
We chose the categories based on our understanding of the complexity of different operations; they are: \code{struct} (for ``structural'' operations \code{extract}, \code{concat}, \code{zero\_extend}, and \code{sign\_extend}), \code{bitwise} (for bitwise operations like \code{bvand}), \code{shift} (for shift operations like \code{bvshl}), \code{linear} (for linear arithmetic operations like \code{bvadd}), \code{mul} (for multiplication), \code{divmod} (for division, modulus, and remainder), and \code{comp} (for comparisons like \code{bvslt}).

\subsubsection{Benchmark Characterization}\label{sec:benchmark_character}

To demonstrate that the \Ozero, \Otwo, and CHC-COMP benchmarks differ according to these features,
we train a multinomial logistic regression classifier to use the features to distinguish between the benchmark categories.
We chose logistic regression because it is interpretable, and---being a relatively weak form of model---less likely to overfit the training data.
To improve the interpretability of the classifier, we train the classifier on normalized features: we take the logarithm of absolute counts, and then z-score all features.
We also remove highly correlated features by iteratively dropping the feature with the highest variation inflation factor (VIF), until every feature has a VIF less than 10.
This results in nine absolute count features being dropped.

The trained classifier successfully distinguishes between five categories of benchmarks, i.e., CHC-COMP and four types of binaries: \Ozero (SV-COMP), \Ozero (other), \Otwo (SV-COMP), and \Otwo (other).
In five-fold cross validation, the classifier achieves a mean accuracy and mean F1-macro score of 0.91 ($\pm$ 0.07).
The classifier's high F1-macro score---the unweighted average of the per-class F1 scores---indicates strong precision and recall for all the benchmark categories.

\begin{figure}[!t]
\centering
\includegraphics[width=1.0\textwidth]{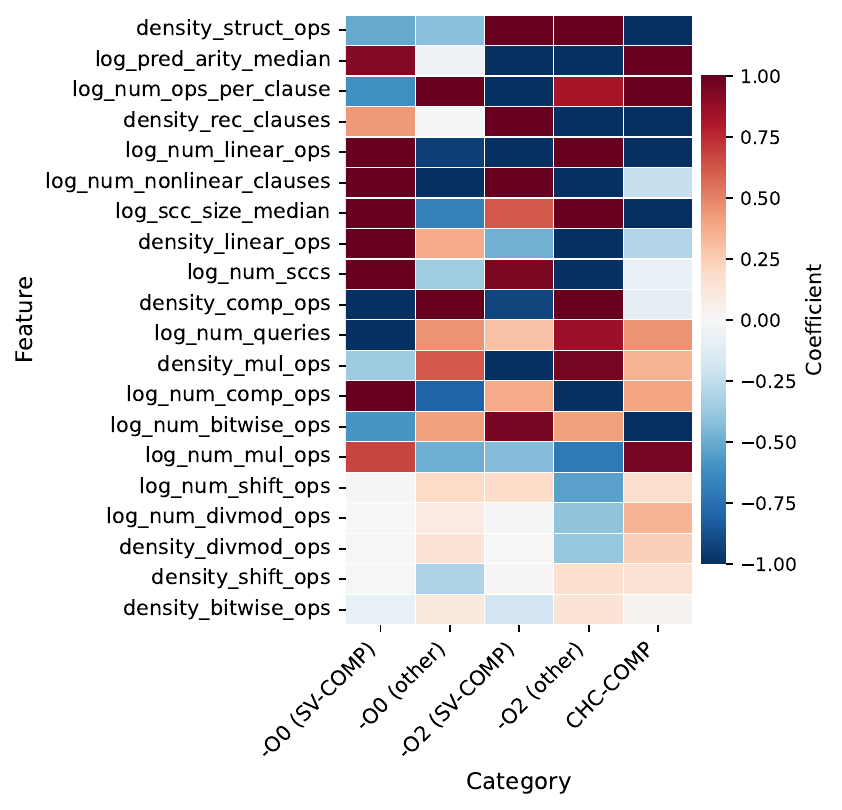}
\caption{Each category of bit-vector CHC problems has a unique ``fingerprint''. Features are ranked by their discriminative power. Positive coefficients indicate the positive association of a feature with a category (the opposite for negative).}
\label{fig:feature-heatmap}
\end{figure}

The five features with the highest discriminative power for the classifier are the density of structural operations, the median predicate arity, the number of operations per clause, the density of recursive clauses, and the number of linear (addition) operations.
For example, the density of structural operations (like \code{extract} and \code{concat}) can differentiate \Otwo benchmarks (high density) from CHC-COMP benchmarks (low density); however, this feature is less useful for classifying \Ozero benchmarks.
The least discriminative features have to do with bit-vector operations that are not well represented in the benchmark sets (e.g., shifts).
Taken together, the coefficients of the multinomial logistic regression for each feature establish the ``fingerprint'' of each benchmark category~(\figurename~\ref{fig:feature-heatmap}).

\begin{tcolorbox}[rqanswerbox]
    \textbf{RQ4 Answer:} Binary-derived benchmarks are structurally different from CHC-COMP benchmarks, and from each other (based on binary type).
\end{tcolorbox}

\begin{figure}[!t]
    \centering
    \includegraphics[width=\textwidth]{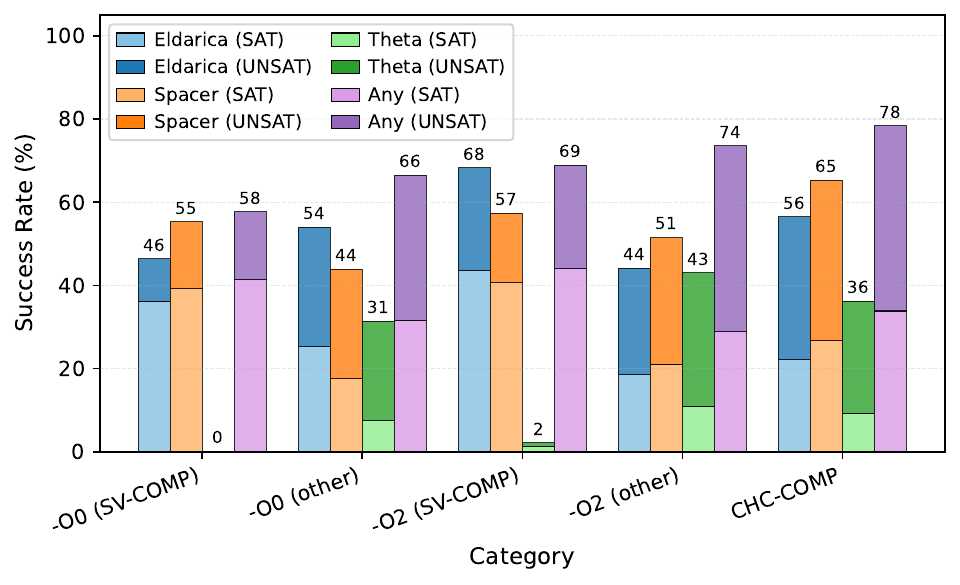}
    \caption{Solver success rates and rankings vary based on benchmark category, suggesting that binary-derived benchmarks can provide meaningful diversity.}\label{fig:solver_comparison}
\end{figure}

\subsection{Performance Diversity}\label{sec:perf_diversity}

In this subsection, we show that the success rates of the solvers and the rankings between solvers vary by 
benchmark category (\figurename~\ref{fig:solver_comparison}).
Following CHC-COMP conventions, we exclude benchmarks where the solvers return inconsistent results, i.e., both \sat and \unsat.
CHC-COMP is the easiest benchmark category, with \SI{78.4}{\percent} of the benchmarks solved by some solver.
In comparison, only \SI{57.4}{\percent} of the \Ozero SV-COMP benchmarks were solved by any solver.

Spacer and Eldarica trade-off on being the best individual solver, depending on category.
Spacer is the most effective solver on the CHC-COMP benchmarks and on two of the binary benchmark categories---the \Ozero (SV-COMP) and \Otwo (other) benchmarks---with an advantage over Eldarica of up to \SI{8.9}{\pp}.
However, Eldarica is the best solver on the \Ozero (other) and \Otwo (SV-COMP) benchmarks, with a \SI{10}{\pp} higher success rate in each case.
It is interesting that no solver dominates a dimension that cuts across categories, such as compiler optimization level, or benchmark provenance (i.e., SV-COMP or other). 

While Theta is consistently the third-ranked solver, its relative competitiveness varies substantially: it is just \SI{1.0}{\pp} less successful than Eldarica on the \Otwo (other) category, yet it solves only \SI{1.2}{\percent} of the SV-COMP problems. 

The effectiveness of portfolio solving also varies greatly by category.
In particular, portfolio solving (using all three solvers) provides little benefit on the SV-COMP benchmarks, having an advantage over the best individual solver of just \SI{2.3}{\pp} and \SI{0.7}{\pp} on the \Ozero and \Otwo SV-COMP benchmarks, respectively.
However, the portfolio solver has an advantage of \SI{22.5}{\pp}, \SI{21.6}{\pp}, and \SI{13.1}{\pp} on the \Ozero (other), \Otwo (other), and CHC-COMP problems, respectively.
This advantage is partly driven by Theta's improved performance here.

\begin{tcolorbox}[rqanswerbox]
    \textbf{RQ5 Answer:}  
    Solver success rates and solver rankings vary among CHC-COMP problems and different types of binary-derived benchmarks.
\end{tcolorbox}

\section{Related Work}\label{sec:related}

\subsection{Automated Binary Analysis}

As surveyed by Kim et al.~\cite{KimFJJOLC17}, automated binary analysis frameworks exist for a range of analysis purposes~\cite{Shoshitaishvili2016SOK,bru11,gid19,mcsema22,KinderV08,2015-djoudi-tacas}.
However, the focus is typically on vulnerability detection, bug finding, and reverse engineering, not verification.
Notable exceptions are CBAT and TINA.
CBAT~\cite{Fortuna2024CBAT} is an extension of the binary platform BAP \cite{bru11} that uses SMT solving to check
the correctness of a transformed intermediate representation relative to the original version. However, CBAT can fail to find some violations due to
approximations like the bounded unrolling of loops.
Adding CHC solving to CBAT could be a fruitful enhancement of that tool.
TINA~\cite{2019-recoules-ase} decompiles assembly code embedded in C code, with the intention that the resulting program can be verified with Frama-C~\cite{kirchner2015frama}.
The decompilation process applies multiple transforms to enhance the verifiability of the code; a translation validation approach ensures correctness.

\subsection{Bit-Vector Reasoning in Automated Verification}

Bit-vector reasoning has long been recognized as a bottleneck to automated verification.
To avoid bit-vector reasoning, some tools~\cite{Lal2014Powering,He2017Counterexample} use mathematical integers to the greatest extent possible, and treat select bit-vector operations imprecisely, as uninterpreted functions.
CHC solvers do not support arbitrary uninterpreted functions (although uninterpreted functions can be simulated using algebraic data types~\cite{Alt2022SolCMC}, which some solvers do support).
Other work encodes bit-vector SMT constraints precisely in an extended theory of
integers~\cite{Zohar2022Bit,Barth2024Bit}.
Much of this prior work assumes that, in the common case, integers are good proxies for bit vectors, which holds true for C verification benchmarks where bitwise operations are rare~\cite{Schuessele2024Ultimate}, but not necessarily for lifted binaries, where bit-vector operations are ubiquitous.
An exception is DarkSea~\cite{Liu2021Proving}, which evaluates its technique---rewriting bitwise operations into conditional expressions over simpler operations---on C programs that are decompiled from binaries and thus contain more bitwise operations.
In their evaluation, DarkSea's technique leads to significant performance improvements when \ultimate is used as the backend verifier.
Newer versions of \ultimate---including the one we evaluated for source-level reasoning---have adopted a generalization of DarkSea's strategy~\cite{Schuessele2024Ultimate}.

\subsection{Instance Characterization for Automated Reasoning}

We are not aware of prior work characterizing CHC problems using features.
However, a common reason to characterize automated reasoning problems by features is for algorithm selection.
SATzilla~\cite{Xu2008SATzilla} chooses which SAT solver to use based on 48 features of a SAT instance; beyond static features (similar to the structural features we discuss in Section~\ref{sec:struct_diversity}), SATzilla also uses dynamic features extracted from short runs of solvers on the problem instance~\cite{Nudelman2004Understanding}.
MachSMT~\cite{Scott2023Algorithm} chooses which SMT solver to use based on 196 static features of an SMT instance (including some for the frequency of individual bit-vector constructs).
Both SATzilla and MachSMT have performed well in solver competitions.

Instance Space Analysis (ISA)~\cite{SmithMiles2023Instance} is a general framework for identifying the classes of benchmarks where a particular solver performs well and evaluating the diversity of benchmark suites.
It projects problem instances from their high-dimensional feature space onto a two-dimensional plane, revealing the ``footprint'' of each solver (where it performs well), as well as gaps in the instance space that are not represented in the current benchmark set.
Applying ISA would be the next logical step in our work on characterizing bit-vector CHC problems.

Despite the recognition that features can usefully characterize benchmarks, competitions related to verification---such as SV-COMP~\cite{Beyer2025Improvements}, SMT-COMP~\cite{SMTCOMP2025Rules}, and CHC-COMP~\cite{DeAngelis2023CHC}---do not, to our knowledge, use features to select diverse benchmark sets.
In these competitions, benchmarks are provided by the external community, and the competition organizers do some filtering to select interesting benchmarks---e.g., they might remove benchmarks that are empirically easy to solve while ensuring that each benchmark source is represented in the final set.

\section{Limitations and Future Work}\label{sec:limitations}

Our results indicate that CHC solving can be a useful component of binary analysis for the binaries we considered---i.e., AArch64 Linux binaries derived from C verification benchmarks used to evaluate invariant inference.
Because invariant inference is a challenging problem in itself, the source programs omit complicating language features 
such as pointers and dynamic memory allocation.
These features require memory to be explicitly represented in the lifted binary.
As \basil represents memory regions with arrays, benchmarks with more complex memory operations would result in CHCs that fall into the logic of quantifier-free bit vectors {\em and bit-vector arrays} (\code{QF\_ABV}).
Spacer, Eldarica, and Theta support this logic, and future work could evaluate the solvers on these CHC problems.

Some rough edges in the bit-vector CHC solving infrastructure could be distorting the performance comparison between solvers.
For example, the CHC-COMP formatting script uses Z3's simplifications, which can result in $n$-ary bit-vector concatenations; Spacer accepts such concatenations (which fall outside the SMT-LIB standard~\cite{Barret2016Satisfiability}), but Eldarica's frontend rejects them.

In our experiments, binary-level reasoning was as successful as source-level reasoning.
This result could be specific to the fact we were looking at programs written in a low-level language, C, which does not have many high-level abstractions: in this context, basic decompilation can close much of the gap between the binary and the source program.
There might be a much larger gap between binary-level and source-level reasoning for higher-level languages, where decompilation is less likely to recover source-level abstractions.
Future work could evaluate CHC solvers on binaries derived from higher-level languages.

We identified 29 static features of bit-vector CHC problems that structurally distinguish different benchmark categories (e.g., CHC-COMP problems vs problems derived from \Ozero SV-COMP binaries).
However, in preliminary experiments using random forest classifiers trained on our benchmarks, the features were not sufficient for algorithm selection or predicting solver success.
These tasks likely require additional features, such as the relevant static features used by MachSMT~\cite{Scott2023Algorithm}, and perhaps dynamic features like those used by SATzilla~\cite{Xu2008SATzilla}.

\section{Conclusions}\label{sec:conclusions}

Our work gives preliminary evidence that bit-vector CHC solving can be a useful component of binary analysis, such as in inferring loop invariants in automated or auto-active verification.
At the same time, we demonstrate that binary analysis is a good source of benchmarks for bit-vector CHC solving, as binary-derived benchmarks are structurally different from---and have different performance characteristics than---the bit-vector benchmarks currently in CHC-COMP.
We have submitted our benchmarks to CHC-COMP 2026, with the hope that the benchmarks will encourage CHC solver developers to continue advancing bit-vector reasoning capabilities in ways beneficial to binary analysis.

\ifmyanon
\else
\begin{credits}

\subsubsection{\ackname}
We thank the anonymous reviewers for their helpful feedback, the Basil developers for answering our questions and fixing issues, and Kirsten Winter, Graeme Smith, and Alicia Michael for many insightful discussions.
The data analysis scripts were created with the help of Claude Code.
This research was supported by The University of Melbourne's Research Computing Services and the Petascale Campus Initiative.
This research is supported by the Commonwealth of Australia as represented by the Defence Science and Technology Group of the Department of Defence.

\subsubsection{\discintname}
The authors have no competing interests to declare that are relevant to the content of this article.

\end{credits}
\fi

\bibliographystyle{splncs04}
\bibliography{main}

\end{document}